\documentclass[aip,jcp]{revtex4-1}
\usepackage{epsfig}

\begin{document}

\title{Comparing parallel and simulated tempering enhanced sampling 
algorithms at phase transition regimes}
\author{Carlos E. Fiore} 
\email{fiore@fisica.ufpr.br}
\author{M. G. E. da Luz}
\email{luz@fisica.ufpr.br}
\affiliation{Departamento de F\'{\i}sica, 
Universidade Federal do Paran\'a, 
CP 19044, 81531-990 Curitiba, Brazil}  
\date{\today}  
 
\begin{abstract} 
Two important enhanced sampling algorithms, simulated (ST) and 
parallel (PT) tempering, are commonly used when ergodic simulations 
may be hard to achieve, e.g, due to a phase space separated by large 
free-energy barriers.  
This is so for systems around first-order phase transitions, a case 
still not fully explored with such approaches in the literature. 
In this contribution we make a comparative study between the PT and 
ST for the Ising (a lattice-gas in the fluid language) and the BEG 
(a lattice-gas with vacancies) models at phase transition regimes.
We show that although the two methods are equivalent in the limit of 
sufficiently long simulations, the PT is more advantageous than the 
ST with respect to all the analysis performed: 
convergence towards the stationarity; 
frequency of tunneling between phases at the coexistence; 
and decay of time-displaced correlation functions of thermodynamic 
quantities.
Qualitative arguments for why one may expect better results from 
the PT than the ST near phase transitions conditions are also 
presented.
\end{abstract} 


\maketitle   

%%%%%%%%%%%%%%%%%%%%%%%%%%%%%%%%%%%%%%%%%%%%%%%%%%%%%%%%%%%%%%%%%%%%%%%%%% 
\section{Introduction}

A keystone procedure to obtain macroscopic thermodynamics quantities 
(e.g., energy, specific heat, magnetization, phase transition points,
etc) of statistical systems is to perform appropriate averages over 
their microscopic configurations.
In practice, however, such systems usually have a prohibitive number 
of states for a full covering.
Therefore, approaches relying on proper representative samplings 
must be considered and so Monte Carlo tools become fundamental for 
calculations.
By a proper sampling we mean that for a given instance a method
should satisfactorily:
(i) represent the way the system actually evolves throughout the 
different microstates (among the whole set $\mathcal{S}$ of
microstates in the system); and 
(ii) generate a set $\Omega$ of visited microstates that indeed 
gives a good picture of all the relevant microstates which 
describe the problem at that particular situation.

Within this framework, an important issue is to know under what
conditions the above criteria are fullfiled. 
For example, biased values for physical quantities may arise when 
the system displays local free-energy minima and the dynamics used 
to generate the microscopic configurations either is not able to 
cross such barriers or it does so, but only after too long times.
Consequently, we have broken ergodicity for finite (even large)
simulations \cite{palmer,neirotti}, leading to metastability and 
thus to poor estimates for the system properties due to a 
non-representative $\Omega$.
Metastability and broken ergodicity appear in several problems like; 
spin-glasses; protein folding, biomolecules; and random search, to 
name just a few \cite{examples-metastability}. 
Moreover, they are not restricted only to complex systems, also 
being present in simpler contexts like in lattice-gas models 
displaying first-order phase transitions 
\cite{cluster1,cluster2,fiore8}. 
As noted, in such case the sampling dynamics may present
difficulties to cross the energetic barriers.
Then, the system can develop hysteresis by passing back 
and forth the phase frontiers as we change the parameter control 
\cite{fiore8}.

Different alternative ideas have been considered to overcome \cite{wang} 
or even circumvent \cite{cluster1,cluster2} entropic barriers, thus 
restoring the ergodic behavior.
In particular, enhanced sampling algorithms, such as parallel tempering 
(PT) \cite{nemoto, geyer, hansmann1} 
-- also known as multiple replica exchange -- 
and simulated tempering (ST) \cite{marinari, pande}, have recently 
attracted a lot of attention, specially due to their simplicity and 
generality compared to other Monte Carlo algorithms 
\cite{cluster1, fiore8}.
Briefly, in the PT method, microscopic configurations in higher 
temperatures are used to assure an ergodic free walk in lower 
temperatures: one simulates replicas of the same system at distinct
$T$'s, allowing the exchange of temperature between the replicas.
For the ST, on the other hand, an unique replica is considered, 
however, the system occasionally undergoes temperature changes 
along its evolution.

Given the different tempering implementation in the two approaches, 
a natural question is how they compare to each other 
\cite{park2, ma, pande2}.
For example, the rate of temperatures switching is higher for the 
ST \cite{park2, ma, pande2}.
So, usually one could expect a larger number of distinct phase space 
regions visited when using the ST, thus a possible advantage over 
the PT.
But as we discuss in Section II.C, near phase transition conditions
this is not always the case.
Therefore, it still an open query if indeed one method is 
systematically superior in all situations.
With the above in mind, here we compare the PT and ST efficiencies 
when applied to phase transitions, specially to the first order 
case.

A short comment regarding the comparison between the PT and ST for 
first order phase transitions is in order.
In principle, for a true first order transition, i.e., for systems 
in the thermodynamic limit, the energy descontinuous gap would lead 
to a small probability of accepting exchanges between the PT replicas 
\cite{nemoto}.
But in concrete calculations, one is always dealing with finite sizes 
$L$, where the actual thermodynamics properties are described by 
continuous functions.
Also, these functions are smooth and tend to the correct asymptotic 
behavior (for $L \rightarrow \infty$) only if the state space is 
properly sampled \cite{cluster2,wang}, what has been shown to be 
the case for the PT \cite{fiore8}.
Thus, in practice the above mentioned difficulty for the PT is not 
an issue and the method is indeed an appropriate tool to study 
first order transitions, as discussed and exemplified in 
different works \cite{fiore8,pt-first-order,hansmann2}.
Hence, the PT and ST (this latter rarely considered in such regime, 
few exceptions being Refs. \cite{st-first-order}) can be analyzed 
at the same footing.
So, possible convergence differences can be associated just to the 
way the algorithms generate the sets $\Omega$, and not to the
approaches eventual instrinsic distinctions (recall that conceptually
they are similar \cite{berg}).

In this contribution we first revisit the simplest Ising spin model 
displaying a well understood second order phase transition.
This is an instructive example because in a recent work \cite{ma}, 
it has been shown that through an improved version of the ST, the 
frequency of successful exchanges (measured in terms of transition
decay rates) is higher for the ST than for the PT method.
However, the comparison was not carried near the critical temperature.
By analyzing time correlation functions, defined as
\begin{equation}
C_{w}(\tau) = 
\langle w(t) - {\bar w} \rangle \langle w(t + \tau) - {\bar w} \rangle,
\label{acf}
\end{equation}
for $w$ relevant thermodynamic quantities (like energy and magnetization) 
of mean $\bar w$ and $\langle \ \rangle$ denoting time averages, one no 
longer gets a better performance of the ST around $T_c$.
In fact, we find that the PT leads to faster decaying $C$'s.

Then, we move to the main focus of this contribution: the harder 
situation of strong first-order phase transitions, where the use
of one-flip algorithms like Metropolis often gives rise to poor 
numerical simulations.
As the specific case study, we consider the lattice gas model with 
vacancies (a spin-1 model in the magnetic systems jargon) 
\cite{BEGMODEL}.
This class of problems has been extensively studied under different 
alternative methods \cite{cluster1,pla,cluster2,fiore8, fiore4}.
Hence, the many available results can help to benchmark those obtained 
from the PT and ST.
We show that although both, PT and ST, lead to equivalent good results 
in the limit of long simulations, the PT displays a faster convergence 
towards stationarity. 
Moreover, for the PT, the tunneling between different phases at the 
coexistence is more frequent and the generated microscopic 
configurations uncorrelate faster.

%A technical note: a fundamental ingredient in the ST is the appropriate 
%definition of weights $g$ \cite{marinari} for the accepting probabilities. 
%Recently, a simple procedure for their estimation has been devised 
%\cite{pande}.
%In this paper we consider such implementation, but we also propose 
%an exact method for calculating the $g$'s. 
%It consists in determining, directly from Monte Carlo simulations,
%the free energies (which are proportional to the $g$'s) by means of 
%the transfer matrix \cite{sauerwein} largest eigenvalue.

The work is organized as the following.
In Sec. II we review the PT and ST methods, discussing distinct 
implementations.
We also give reasons why the PT may outperform ST near phase 
transition conditions.
In Sec. III we consider a spin system displaying a second-order 
phase transition.
The lattice-gas model and its comparative study with the PT and 
ST methods -- addressing a first order phase transition -- are 
presented in Sec. IV.
Finally, in Sec. V we draw our last remarks and the conclusion.

\section{The PT and ST sampling algorithms}

The central idea behind a tempering enhanced sampling algorithm
is try to guarantee ergodicity by means of appropriate temperature 
changes during the simulations, thus allowing efficient and uniform
visits to a fragmented multiple regions phase space \cite{berg}.
Suppose we shall study a system at a given $T_0$.
We assume $T_1 = T_0$ and define a set of $N$ distinct temperatures 
$T_1 < T_2 < \ldots < T_N$, with $\Delta T = T_N - T_1$.
There are different ways to implement tempering 
\cite{tempering-implementation}, two important ones being 
the PT and ST, which we describe next.

%
%For the PT, $N$ replicas of the system -- at the different 
%$T_n$'s -- evolve according to a given prescription and can have 
%their temperatures interchanged along the way.
%For the ST, the temperature becomes a dynamical variable too.
%Thus, besides the dynamical prescription applied to a single 
%copy of the system, $T$ also assumes the distinct values $T_n$'s 
%along the simulations. 
%Next we describe each method in more details.

\subsection{Parallel Tempering}

The PT approach combines a standard algorithm (e.g., Metropolis)
with the simultaneous evolution of $N$ copies of the system
(each at a different $T_n$), occasionally allowing the replicas to 
exchange their temperatures. 
Fixing relevant parameters, the method is implemented by first 
running $M_{eq}$ times (to assure equilibration of all the $N$ 
copies) a two parts procedure, (a) and (b), discussed below.
After that, for each (a)-(b) composite MC step (repeated  $M_{a,b}$ 
times) we calculate the thermodynamics quantities at the 
temperature of interest $T = T_1$.
The average over the $M_{a,b}$ partial values give the final 
results.
In fact, we further improve the calculations and estimate the 
statistical deviations by performing this procedure (after 
relaxation) $M_{rep}$ times, so that in total the number of (a)-(b) 
MC steps is $M_{tot} = M_{eq} + M_{a,b} \times M_{rep}$.

In (a), for each replica (at a distinct $T_n$), a site lattice $l$ 
is chosen randomly. 
Then, its occupation variable $\sigma_l$ may change to a new value 
$\sigma_{l}'$ according to the Metropolis prescription 
$P = \rm min\{1, \, \exp[-\beta \Delta \cal H]\}$ \cite{metr}, where
$\Delta{\cal H} = {\cal H}(\sigma') - {\cal H}(\sigma)$ is the 
energy variation due to the occupation change.
This is done until a full lattice covering and the process is 
repeated all over again $M$ times. 
(b) In the second part, arbitrary pairs of replicas (say, at $T_{n'}$ 
and $T_{n''}$ and with microscopy configurations $\sigma'$ and 
$\sigma''$) can undergo temperatures switchings, with probability 
($\beta_n = (k_{B} T_{n})^{-1}$)
\begin{equation}
p_{n' \leftrightarrow n''} = 
\min \{1, \, \exp[(\beta_{n'} - \beta_{n''})
({\cal H}(\sigma') - {\cal H}(\sigma''))] \}.
\label{p-pt}
\end{equation}
The PT algorithm is schematic represented in Fig. 1 (a).

Although the above prescription is rather simple, few technical 
aspects should be observed.
First, it is necessary to find a good compromise between the $p$'s
values (which increase with $\Delta T/N$ decreasing) and the replicas 
number $N$.
This is so to guarantee relatively frequent exchanges, while keeping 
the computational efforts low.
Hence, extra procedures have been proposed 
\cite{hansmann2,helmut,predescu,doll,bittner}.
Here we use only the ones explained above.
However we mention that for our present systems, one of us has tested 
some of these extra implementations \cite{fiore8} (always assuming 
arbitrary $n'$'s and $n''$'s for the step (b) above), not finding any 
significant difference. 
Second, the system size ($L$) also imposes restrictions on the $N$'s.
For small systems, a few number of replicas is enough to assure rapid 
convergence.
On the other hand, by increasing $L$ the exchange probabilities
(Eq. (\ref{p-pt})) decreases, so the inclusion of extra copies 
becomes necessary.
Such care has been explicit taken in our simulations.
Finally, we observe that most works that use the PT method implement 
the switching attempts only between adjacent replicas (i.e., at 
$T_{n'}$ and $T_{n'' = n'+1}$), in principle because the probability of 
exchanges decreases for increasing $T_{n''} - T_{n'}$.
Nevertheless, it has been shown \cite{fiore8} that non-adjacent 
exchanges are essential to speed up the crossing of high free-energy 
barriers (what we discuss in more details in Section II.C).
Therefore, here we will allow exchanges between first ($\delta = 1$), 
second ($\delta = 2$), etc, neighbor replicas, meaning those 
between $T_n$ and $T_{n + \delta}$.

\begin{figure}
\setlength{\unitlength}{1.0cm}
\includegraphics[scale=0.45]{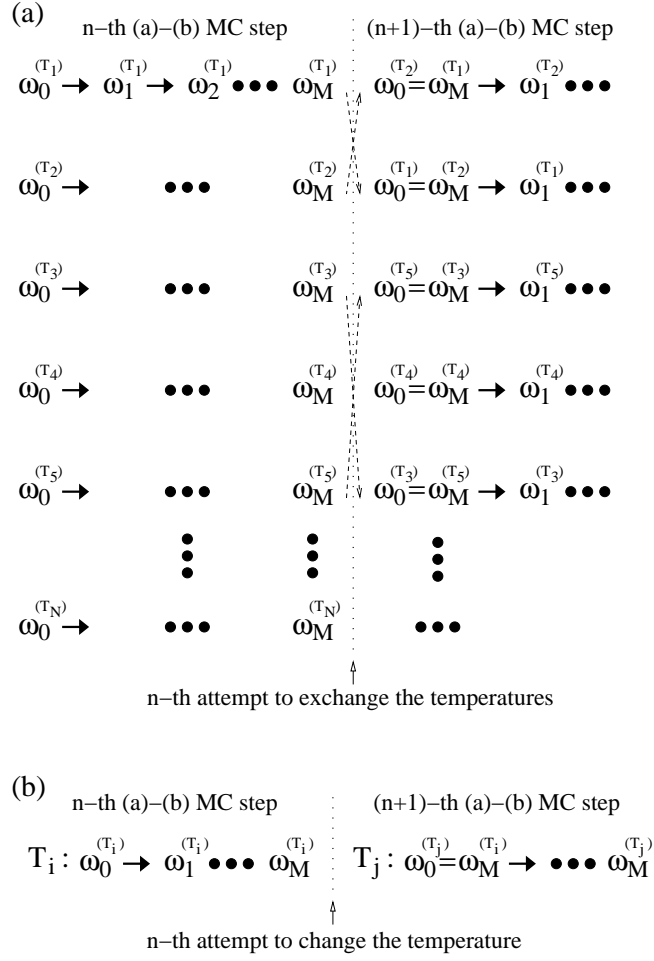}
\caption{Schematics of the (a) PT and (b) ST implementations. 
In this example, there have been two temperatures exchanges for 
the PT ($T_1 \leftrightarrow T_2$ and $T_3 \leftrightarrow T_5$)
and one temperature change for the ST ($T_i \rightarrow T_j$).}
\end{figure}

\subsection{Simulated Tempering}

For the ST, a single realization of the model is considered,
however, during the dynamics its temperature can assume the
different values $T_n$'s.
The implementation is similar to that for the PT in Section
II.A, but applied only to one copy of the system.
Therefore, the previous step (b) now reads:
A change $T_{n'} \rightarrow T_{n''}$ may take place for the
system according to the probability (with $\sigma$ its 
configuration)
\begin{equation}
p_{n' \rightarrow n''} = 
\min \{ 1, \, \exp[(\beta_{n'} - \beta_{n''}){\cal H}(\sigma) 
+ (g_{n''} - g_{n'})] \}.
\end{equation}
The ST algorithm is illustrated in Fig. 1 (b).

Note that $p_{n' \rightarrow n''}$ depends on the weights $g$'s.
Moreover, for a better sampling, the evolution should uniformly 
visit all the established temperatures.
This is just the case when $g_{n} = \beta_{n} \, f_{n}$, with $f_n$ 
the system free energy at $T_n$ \cite{pande,park2,pande2}. 
To obtain $f$ is not an easy task.
For instance, in Ref. \cite{ma} its exact (numerical) values 
follows from $f_n = - \ln[Z_n] / (V \beta_n)$, with the partition 
function $Z_n$ computed by an involving recursive procedure.
Here, $V$ is the system volume, which in a regular square lattice
reads $V = L^2$.
In our examples we will consider this same protocol, but using a
simpler numerical implimentation for $Z_n$.
Indeed, in the thermodynamic limit 
\begin{equation} 
Z_n = - (\lambda^{(0)}_n)^L,
\label{e4}
\end{equation}
where $\lambda^{(0)}_n$ is the largest eigenvalue of the transfer 
matrix ${\mathcal T}$ at $T_n$ (for details see, e.g., Ref. 
\cite{sauerwein}).
By its turn,  
$\lambda^{(0)} = \langle {{\mathcal T}(S_{k}, S_{k})} \rangle /
\langle \delta_{S_{k}, S_{k+1}} \rangle$ can be calculated from 
straightforward Monte Carlo simulations \cite{sauerwein}, where
$S_{k}$ is the lattice $k$-layer configuration
$\sigma_{1,k}, \sigma_{2,k}, \ldots, \sigma_{L,k}$ and
$\delta_{S_{k}, S_{k+1}} = 1$ ($= 0$) if the $k$ and $k+1$ layers
are equal (different).
A central point is that in principle Eq. (\ref{e4}) would hold true 
only for infinite size systems.
However, if $L$ is not too small, the above relation is extremely 
accurate and for any practical purpose gives the correct $Z_n$,
as we show in the next Section.
Such way to determine $p_{n' \rightarrow n''}$ will be named the ST 
(exact) free-energy method, ST-FEM.

Finally, we observe that approximations for $g$ are equally possible.
One implementation being \cite{pande} 
\begin{equation}
g_{n+1} - g_{n} \approx 
(\beta_{n+1} - \beta_{n}) (U_{n+1}+U_{n})/2,
\end{equation} 
with $U_n = \langle {\cal H}_n \rangle$ ($n=1, 2, \ldots, N$)
the average energy at $T_n$.
The $U$'s can be evaluated from direct auxiliary simulations. 
For completeness we will also consider this ST approximated method, 
which we call ST-AM.

%Finally, to make sure that the results here should not be credited
%to any artifact of our different numerical procedure for the 
%ST (ST-FEM), we have compared it with an implementation based 
%directly on the numerical calculation of the partition function 
%\cite{zhang, ma}. 
%For instance, we have evaluated the partition function per volume,  
%which is proportional to $\lambda^{(0)}$, from the present approach 
%and compared to those obtained in Ref. \cite{zhang} (results not
%shown), finding complete agreement.

\subsection{The PT and ST methods near phase transition 
regimes}

\begin{figure}
\setlength{\unitlength}{1.0cm}
\includegraphics[scale=0.5]{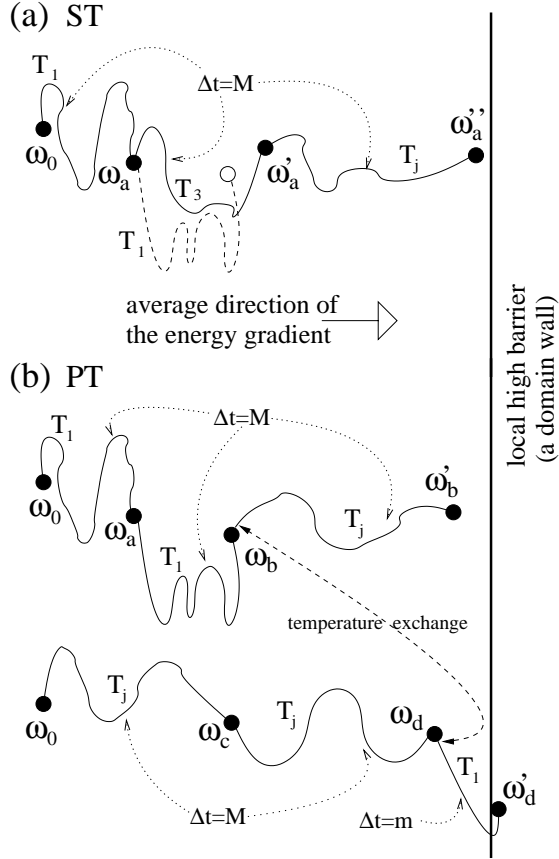}
\caption{Schematic illustration of the trajectories -- succession of 
$\omega$'s -- generated by the PT and ST algorithms in the case 
of a complex topography for the relevant microstate space
(resulting from specific parameters values).
A higher sinuosity (usually associated to smaller $T$'s)
represents a higher difficulty to leave the particular region 
of $\mathcal{S}$, full of energetic valleys and hills.
The length of the paths are proportional to the number of
steps taken by the algorithms.}
\end{figure}

The sampling of a statistical system when the phase space has 
a complicated landscape full of free-energy valleys and hills 
\cite{mauro1} is particularly delicate: one needs to uniformly 
visit different regions of $\mathcal{S}$ \cite{debenedetti} 
(those more important for the given parameters), but which are 
separated by many entropic barriers \cite{helmut}.
In this case, the particular way in which a method evolves throughout 
the microstates space to generate $\Omega$ -- even with the use of 
enhanced procedures -- may crucially determine the final outcome of 
sampling.
For instance, non-ergodic ``probing'' of the multiple domains 
\cite{williams} can prevent the proper relaxation to equilibrium.

The previous comments fit perfectly well first-order phase 
transitions, where the minima of the free-energy are separated by 
large barriers.
Nevertheless, we observe that for second-order phase transitions, the 
divergence of time and spatial length correlations creates strongly 
correlated configurations \cite{landau-binder}.
It leads to a certain clusterization of relevant parts of $\mathcal{S}$ 
at the critical point, with independent and unbiased $\Omega$ 
difficult to obtain.
So, although associated to different mechanisms, near both first 
and second order transitions we can expect a ``fragmented'' phase 
space.
Hence, even if the PT and ST are not crucially distinct in usual 
situations (in fact, the ST being slight better than the PT in few 
instances \cite{ma}), here we argue qualitatively that in such 
cases the PT can outperform the ST.
 
Thus, for the above contexts of multiple basins \cite{mauro2}, the 
Fig. 2 schematically represents ``stretches'' of typical dynamical 
paths generated by the ST and PT algorithms.
The successively visited $\omega$'s until leaving the domain -- 
delimited by high local free-energy barriers (or cluster walls) 
-- can form a very sinuous trajectory on that particular region 
of $\mathcal{S}$ due to a complex topography.

Thus, consider first the ST, Fig. 2 (a). 
The initial microstate $\omega_0$ evolves (at $T = T_1$) in a very
tortuous path, but in average towards the border of the domain,
reaching $\omega_a$ after $M$ steps.
Then, it undergoes a temperature change $T_1 \rightarrow T_3$ and 
again evolves $M$ steps getting to $\omega_a'$, this time in a more 
straight trajectory because the higher $T$ (note if there was no 
temperature change, the path would follow the dashed line displayed 
in the plot).
Finally, there is a second successful attempt to change $T$, 
$T_3 \rightarrow T_j > T_3$, and after $M$ steps the system ends
up very close to the barrier separating the basins.

In Fig. 2 (b) we observe the PT dynamics, where just one successful
temperature exchange takes place (between the only two replicas 
depicted).
The microstate $\omega_b$ ($\omega_d$) is obtained from $\omega_0$ 
after $2 M$ steps at $T = T_1$ ($T = T_j$).
Obviously, $\omega_a'$ in the ST must be in average closer to 
(farther from) the domain border than $\omega_b$ ($\omega_d$) 
in the PT implementation.
Then, there is an exchange of temperatures and the evolution of
$\omega_d$ at $T_1$, after $m < M$ steps, already makes the 
replica to cross the basin barrier to the microstate $\omega_d'$.
Furthermore, after $\Delta t = M$ the state $\omega_b$ at $T_j$ 
leads to a $\omega_b'$ close to the border.

The above illustration -- although certainly not extinguishing
all the possibilities -- is already representative of why the PT 
can be more efficient in sampling a space full of energetic 
valleys and hills (e.g., at phase transition regimes).
It is so for the following reasons:
(i) In the PT, the existence of replicas at all the interval
$\Delta T$ of temperatures generate some paths which more 
quickly will approach the domain borders, as seen in Fig. 2 (b) 
for $\omega_0 \rightarrow \omega_d$ at $T_j$.
Moreover, the microstates along such trajectories at higher $T$'s 
of course are usually more energetic.
(ii) So, when finally there is an exchange of temperature, a
microstate of high energy, even if now at lower $T$'s, will demand 
a smaller number of steps to cross a barrier (like 
$\omega_d \rightarrow \omega_d'$ in Fig. 2 (b)), and thus to start 
visiting other basins.
On the other hand, trajectories of microstates of low energy, that
during a certain $\Delta t$ have evolved under small values of 
$T$'s, e.g. $\omega_0 \rightarrow \omega_b$  in Fig. 2 (b),
when shifting to higher temperatures will speed up their ways 
towards the barrier ($\omega_b \rightarrow \omega_b'$). 
Note, nevertheless, that this is possible {\em only} if non-adjacent 
exchanges are allowed, the case we are assuming here.
(iii) The above collective dynamics makes possible many of the
replicas successfully leave a domain after fairly similar number 
of steps. 
Hence, once in another basin region, this ``parallel'' process 
can proceed in the same fashion.
(iv) By its turn, we can face the ST as a ``serial'' process, 
then a faster drift towards the domain walls takes place only 
when $T$ increases.
As a consequence, the eventual more frequent temperature exchange 
for the ST \cite{park2, ma, pande2} not necessarily constitutes an 
advantage in complex $\mathcal{S}$ landscapes (as illustrated in 
Fig. 2).
(v) Lastly, a not critical issue but which also may give some
small advantage for the PT over the ST is that in the former, 
often the replicas (even at smaller $T$'s) cross the domain 
high barriers more or less at the same time. 
Thus, once leaving a certain basin we already have a sample of 
microstates at $T_1$ to make averages for the PT.
As displayed in the Fig. 2 (a), for the ST it may happen that 
when the system reaches a microstate configuration able to
cross the barrier, it is not at $T_1$.
Hence, an extra time is necessary for the system (naturally from 
the algorithm dynamics) to come back to $T_1$ and so the averages 
to be performed.

We finally observe that when the relevant space is more homogeneous 
in energy (e.g., far away from phase transitions), one should not 
expect so high increase of the trajectories sinuosity as we diminish 
$T$.
Then, it is not difficult to realize that the listed differences
between the PT and ST methods might not be important.

The previous discussion is based on qualitative arguments. 
Of course, they should be corroborated by concrete quantitative 
studies.
Next we analyze two systems near phase transition conditions.
We will explicit show through detailed numerical simulations
that indeed the PT algorithm is more efficient, specially in the 
case of first order phase transitions.

\section{The Ising model}

The model is defined by the following Hamiltonian
\begin{equation}                                                            
{\cal H} = - J \sum_{<i,j>} \sigma_{i} \, \sigma_{j} 
           - H \sum_{i=1}^{V} \sigma_i,       
\end{equation}
where $<i,j>$ denotes nearest-neighbors pairs $i$ and $j$ of a 
$d$-dimensional lattice of $V = L^{d}$ sites. 
At each site $i$, the spin variable assumes the values
$\sigma_i = \pm 1$.
$J$ is the interaction energy and $H$ is the magnetic field. 
The Ising model displays a second-order phase transition 
(ferromagnetic--paramagnetic) at $T_c \approx 2.269$ and 
$H=0$.
For a square lattice ($d=2$), the transfer matrix diagonal
elements are 
\begin{equation}                                                                
{\mathcal T}(S_{k}, S_{k}) = \exp\left[ \beta \, \big(
\sum_{l=1}^{L} J \,
(1 + \sigma_{l,k} \, \sigma_{l+1,k}) + H \, \sigma_{l,k} \big) \right].      
\label{e17}                                                                     
\end{equation}
%\begin{equation}                                                   
%{\mathcal T}(S_{k}, S_{k}) = \exp\left[
%\sum_{l=1}^{L} \beta \, J \, \sigma_{l,k} 
%(\sigma_{l,k+1} + \sigma_{l+1,k}) + H \right].      
%\label{e17}                                                               
%\end{equation}

%
%--------------------- figure ------------------------- 
%
\begin{figure}
\setlength{\unitlength}{1.0cm}
\includegraphics[scale=0.35]{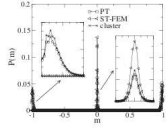}
\caption{For the Ising model with $H=0$, $L=32$,
and units of $J/k_B$, comparison between the partition 
function versus $T$ calculated exactly \cite{ferdinand} 
and from Eq. (\ref{e4}).}
\end{figure}
%
%--------------------- figure -------------------------
%
%
%--------------------- figure ------------------------- 
%
\begin{figure}
\setlength{\unitlength}{1.0cm}
\includegraphics[scale=0.35]{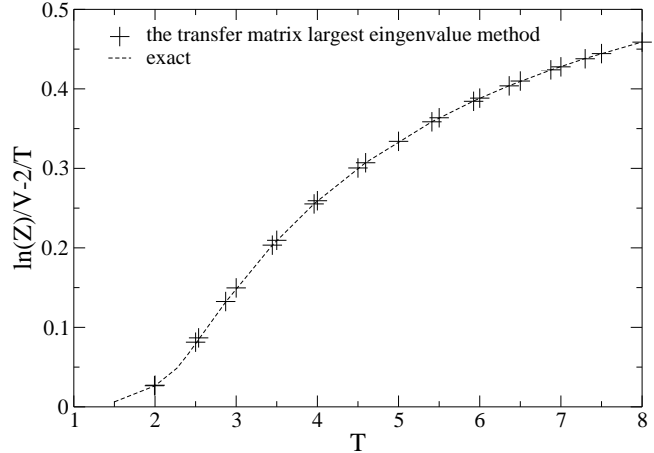}
\caption{For the Ising model at $T_c$,
the auto-correlation functions versus $\tau$ (in MC steps unities), 
simulated from the PT (continuous), ST-FEM (dashed) and ST-AM 
(dotted).}
\end{figure}
%
%--------------------- figure -------------------------
%
%
%--------------------- figure ------------------------- 
%
\begin{figure}
\setlength{\unitlength}{1.0cm}
\includegraphics[scale=0.36]{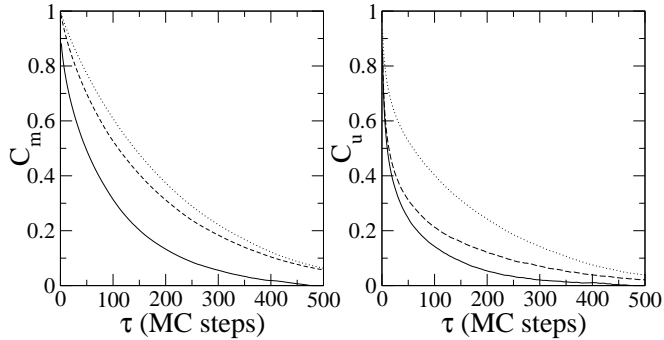}
\caption{For the Ising model at $T_c$,
the time evolution of $u$ and $m$ from a non-typical initial
configuration simulated by the PT (continuous), ST-FEM (dashed)
and ST-AM (dotted).}
\end{figure}
%
%--------------------- figure -------------------------
%

Our interest are in the energy $u = \langle {\cal H}\rangle/V$ 
and modulus of the magnetization (which is the order parameter)
$m = \langle |\sum_{i=1}^{V} \sigma_{i}| \rangle / V$ per volume.
For their auto-correlation functions, we just set $w = u$ and 
$w = m$ in Eq. (\ref{acf}).
Regarding the parameters, we choose $H = 0$ and a square lattice 
of $L = 32$. 
All the results are given in units of $J/k_B$. 
To test the accuracy of the transfer matrix largest eigenvalue
method in obtaing $Z$, in Fig. 3 we compare the exact partition 
function (obtained from the solution in Ref. \cite{ferdinand}) 
with that calculated from Eq. (\ref{e4}) for the Ising model 
and the above parameters.
The agreement is indeed remarkable, indicating that even for 
$L = 32$, $Z$ and consequently $f$ is already very close to the 
thermodynamic limit value.

Figure 4 displays $C_m$ and $C_u$ for $T_1 = T_c$. 
In the simulations we use only two replicas (with $T_2 = 2.4$) 
and $M=1$.
From the plots we see that the auto-correlations decay faster 
when calculated by the PT than by both the ST-AM and ST-FEM
methods.
In Fig. 5 we compare the time evolution of the thermodynamic 
quantities starting from a ``hard'' initial condition, i.e., 
a configuration very different from the ones representative of 
the steady state.
Thus, we consider a fully ordered configuration, which obviously 
is not typical at $T = T_c$.
This is a way of testing how efficient is a certain approach to 
drive the system to the stationary state.
The Ising model at the transition temperature evolves to the 
equilibrium basically in the same fashion either when simulated 
by the PT or by both the ST's.

So, we have that for a continuous phase transition (at least for 
the Ising model) the performances of the two tempering methods are 
essentially equivalent.
Although at $T_c$ the PT shows faster auto-correlation decays
(in contrast with the results of Ref. \cite{ma} for the same
model, however calculated far away from the critical temperature), 
the stationary state is characterized by equivalent values of 
$m$ and $u$ for all methods.

\section{The lattice-gas model with vacancies (BEG)}

\subsection{Model}

The lattice-gas model (of size $V = L^d$) with vacancies is 
characterized by the Hamiltonian
\begin{equation}                                                            
{\cal H} = - \sum_{<i,j>} \sum_{r,s} \epsilon_{r,s} \, N_{r,i} \, N_{s,j} 
- \sum_{r} \sum_i \mu_r \, N_{r,i} .
\label{e1}       
\end{equation} 
Here, $r$ and $s$ run over the species labels $A$ and $B$, the
$\epsilon_{r s}$'s are the coupling energies ($\epsilon_{AA}$, 
$\epsilon_{BB}$, $\epsilon_{AB}$ and $\epsilon_{BA}$), 
$N_{r, i} = 0, 1$ is the occupation numbers at site $i$ for species
$r$, and $\mu_r$ is the species $r$ chemical potential.
The above model is equivalent to the Blume-Emery-Griffiths 
(BEG) spin-1 ${\cal H}$  \cite{BEGMODEL}.
Indeed, defining (with $\sigma_i = 0, \pm 1$ the possible values for 
the spin variable)
\begin{equation}                                                  
N_{A,i} = (\sigma_i^2 + \sigma_i)/2, \qquad 
N_{B,i} = (\sigma_i^2 - \sigma_i)/2,                             
\label{e2}                                                                   
\end{equation}
associating $\sigma_i = 1$ (-1) with the species $A$ ($B$) and 
$\sigma_i = 0$ with a vacancy, and setting 
$\epsilon_{AA} = \epsilon_{BB}$ and $\epsilon_{AB} = \epsilon_{BA}$,
 we get the BEG Hamiltonian
\begin{equation}                           
{\cal H} = 
-\sum_{<i,j>} (J  \, \sigma_{i} \, \sigma_{j} + K \, 
\sigma_{i}^{2} \, \sigma_{j}^{2}) 
- \sum_{i} (H \, \sigma_i - D \, \sigma_i^2),
\label{e3}                                            
\end{equation}
for
\begin{eqnarray}                                                     
& & {\it H} = (\mu_{A} - \mu_{B})/2, \qquad D = - (\mu_{A} + \mu_{B})/2,
\nonumber \\
& & J = (\epsilon_{AA} - \epsilon_{AB})/2, \ \ \ \ 
K = (\epsilon_{AA} + \epsilon_{AB})/2.
\label{e7}                 
\end{eqnarray}

We will consider a square lattice with periodic boundary conditions.
In this case, the transfer matrix diagonal elements read
\begin{eqnarray}                                                
{\mathcal T}( S_{k},S_{k}) &=& 
\exp\Big[ \beta \sum_{l=1}^{L} 
\Big( (H + J \, \sigma_{l+1,k}) \, \sigma_{l,k} \nonumber \\
& &  + (J - D + K (1 + \sigma_{l+1,k}^{2})) \sigma_{l,k}^{2} 
\Big) \Big].
\label{e18}
\end{eqnarray}
%\begin{eqnarray}                                                
%{\mathcal T}( S_{k},S_{k+1}) &=& 
%\exp\Big[ \beta \sum_{l=1}^{L} 
%\Big( J (\sigma_{l,k+1} + \sigma_{l+1,k}) \sigma_{l,k} \nonumber \\
%& &  + K (\sigma_{l,k+1}^{2} + \sigma_{l+1,k}^{2}) \sigma_{l,k}^{2}
%\nonumber \\
%& & - D \sigma_{l,k}^{2} + H \sigma_{l,k} \Big) \Big].      
%\label{e18}               
%\end{eqnarray}

The model has two order parameters, $q$ and $m$, defined by 
$q = \langle \sum_{i=1}^{V}(N_{A,i}+N_{B,i} ) \rangle/V$ and
$m = \langle \sum_{i=1}^{V} (N_{A,i}-N_{B,i}) \rangle/V$.
Also important is the quantity energy per volume, given by 
$u = \langle {\cal H} \rangle/V$.
The auto-correlation are then obtained from $w = q$, $w = m$ 
and $w = u$ in Eq. (\ref{acf}).

\subsection{Results}

For fixed $K/J$, $H$ and $T$, the characteristic of the phase 
space is determined by $D$.
In the regime we are interested, there are two phases if $D$ is small, 
one rich in species A and the other in species B.
For high values of $D$, the model displays a single gas phase, 
rich in vacancies. 
A strong first-order phase transition between these two situations
takes place at $D = D^{*}$, which obviously depends on $K/J$, $H$ 
and $T$.
For definiteness, in the following we study the BEG Hamiltonian 
assuming $K/J = 3$, $H = 0$ and $T = T_1 = 1.4$ (for other parameter 
values, see Sec. V).
In such case, $D^{*} = 8.000$ in the thermodynamic limit 
\cite{fiore8}.
All the results will be presented in units of $J/k_B$.

It is well known that for different lattice-gas systems, approaches 
based on cluster algorithms \cite{cluster1} are very appropriate to 
deal with metastability arising in first-order phase transitions. 
So, next we will compare results obtained from both tempering methods
with those available from cluster calculations \cite{cluster1}.
Regarding the parameters values, unless otherwise explicit mentioned, 
in the simulations we consider $L = 20$, $D = 8.000$ and the replicas 
in the temperature interval $\Delta T = 0.6$.
Also, whenever necessary we perform in total up to 
$M_{tot} = 8 \times 10^{7}$ simulation steps (see Sec. II.A)
to evaluate the sought quantities.
Furthermore, we always use $M=1$.

%We note that as previously mentioned, standard algorithms 
%usually are not able to cross high free-energy barriers, consequently 
%large hysteresis emerges at the phase coexistence.
%An efficient enhanced sampling algorithm is just one that can
%overcome such difficulties.
%Therefore, by properly tuning $D$ to the values corresponding to
%a phase transition, we can test the efficiency of both tempering 
%methods.

As the first comparative analysis, in Fig. 5 we plot the order
parameter $q$ probability distribution histogram for a long 
simulation run of $10^{7}$ MC steps.
As the chemical potential we set $D = 8.004$, instead of 
$D = 8.000$, since it leads to a same high for the two peaks of 
the bimodal order parameter probability distribution 
(we mention, nevertheless, that $D = 8.000$ gives the same 
qualitative results).
The agreement of the two tempering with the cluster method 
\cite{cluster2} is similar (in fact, a little better for the PT case). 
Such calculations show that for a long enough time, both the PT and
ST are able to circumvent the metastable states, allowing the system 
to cross the free-energy barriers separating the different phases 
at the coexistence. 

%
%--------------------- figure -------------------------                        
%
\begin{figure}
\setlength{\unitlength}{1.0cm}
\includegraphics[scale=0.36]{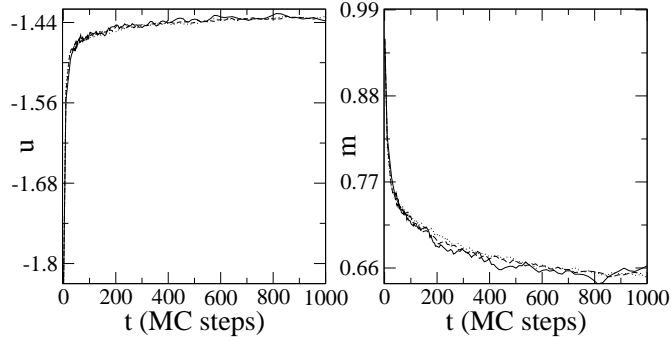}                                       
\caption{For the BEG model, the histograms of the order parameter 
$q$ from a long simulation using the PT, ST and cluster 
algorithms.
The insets are blow-ups of the (a) low and (b) high densities 
regions, $q \approx 0$ and $q \approx 1$, respectively.}
\end{figure}
%
%------------------------------------------------------            
%
%
%--------------------- figure -------------------------                        
%  
\begin{figure}
\setlength{\unitlength}{1.0cm}
\includegraphics[scale=0.35]{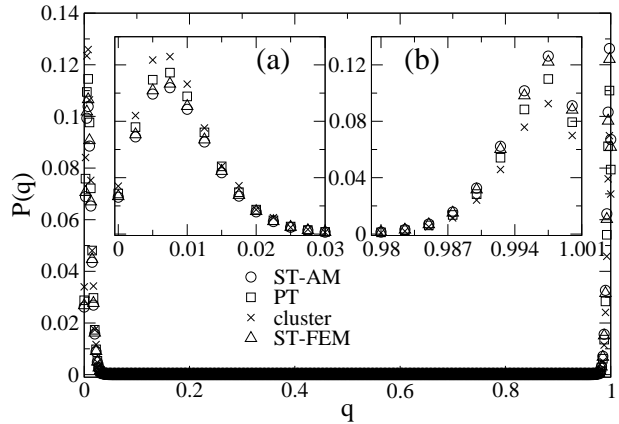}
\caption{For the BEG model, the time evolution of $q$ from
a fully random initial configuration, simulated from the PT, 
ST, and cluster.
$N$ denotes the number of replicas and $\Delta T = 0.6$ if not 
otherwise specified in the curves.}
\end{figure}
%
%--------------------- figure -------------------------                        
%

Despite the previous agreement, the PT and ST do present 
differences when other aspects are analyzed.
For instance, we show in Fig. 6 the time evolution of 
$q$ towards the steady state, starting from a fully random initial 
configuration.
We also consider distinct number of replicas $N$ and temperature
intervals $\Delta T $.
We find that under the same simulation conditions, generally
the PT converges faster, being closer to the cluster results 
than the ST (ST-FEM and ST-AM).
However, for the lower value of $\Delta T = 0.25$, in all cases the 
system (up to $10^4$ MC steps) cannot even escape the region near 
the initial random configuration. 
On the other hand, by increasing $\Delta T = 0.6$ -- although
the probability for temperature exchanging decreases -- the system 
starts to move towards the stationary regime.  
Furthermore, the larger the number of replicas $N$, the faster 
the convergence.
Finally we mention that the steady value of $q = 2/3$ at 
$D = D^{*} = 8.000$ can be understood recalling that at the 
phase coexistence, two liquid phases ($q \approx 1$) coexist 
with one gas phase ($q \approx 0$). 
Since their weights are equal (1/3), we have $q \approx 2/3$ for 
any system size.

%
%--------------------- figure -------------------------                        
%
\begin{figure}
\setlength{\unitlength}{1.0cm}
\includegraphics[scale=0.5]{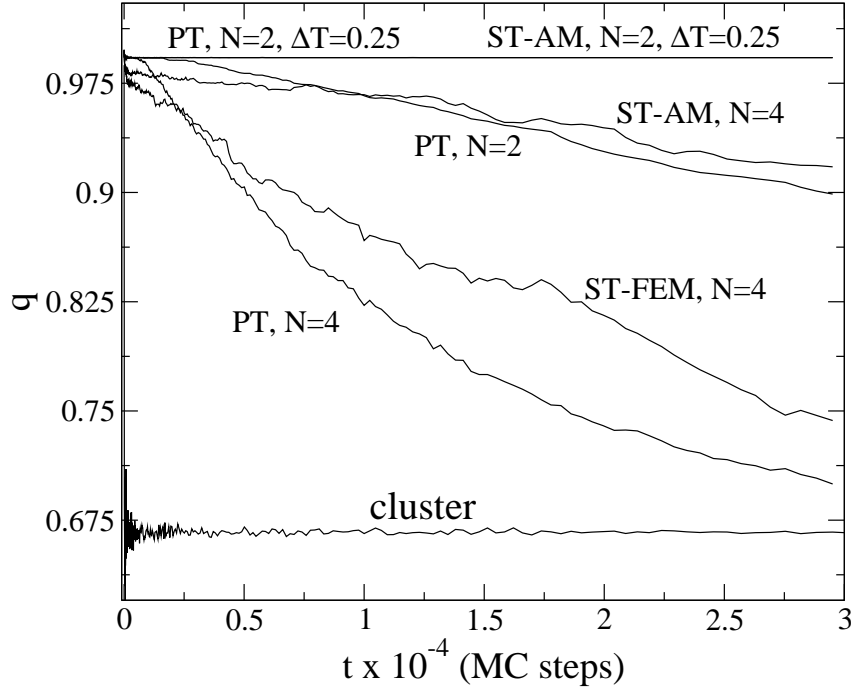}
\caption{For the BEG model, $m$ versus $t$ in two distinct time 
intervals at the steady state (after $M_{eq}$), calculated with the 
PT, ST-FEM and ST-AM algorithm.}
\end{figure}
%
%--------------------- figure -------------------------                        
%
%
%--------------------- figure -------------------------                        
%
\begin{figure}
\setlength{\unitlength}{1.0cm}
\includegraphics[scale=0.38]{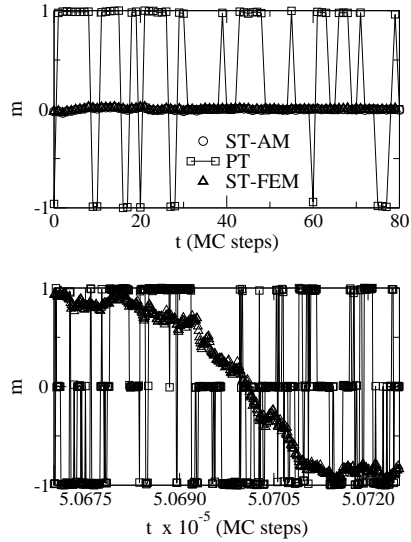}
\caption{Similar to Fig. 7, but comparing PT and cluster.}
\end{figure}
%
%--------------------- figure -------------------------                        
%
%
%--------------------- figure -------------------------                        
%
\begin{figure}
\setlength{\unitlength}{1.0cm}
\includegraphics[scale=0.3]{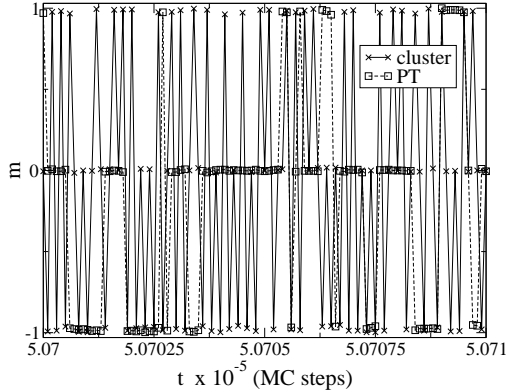}
\caption{For the BEG model, $q$ versus the chemical potential, 
simulated from the PT (square), ST-FEM (triangle), ST-AM (circle), 
and cluster ($\times$).
The averages are taken at each $10^{4}$ MC steps.
In the inset, exactly the same curve but for the averages at each 
$5 \times 10^{4}$ MC steps.}
\end{figure}
%
%--------------------- figure -------------------------                        
%

Another interesting test is to perform the numerical 
simulations when the system is already at the steady state. 
In Fig. 7 we show the time evolution of the ``magnetization'' 
$m$ for both tempering methods at the phase coexistence.
In the plots the time is shifted so to discard the $M_{eq}$
initial MC steps necessary for equilibration.
We see that the tunneling between the three different phases is 
substantially more frequent for the PT than for the ST.
It being true along the whole evolution, as we have checked 
for an interval of $10^{7}$ MC steps (in the Fig. 7 we show only two 
distinct simulation stretches).
Actually, the PT tunneling pattern presents the same behavior than 
that observed in the notorious accurate cluster algorithm 
\cite{cluster1}, Fig. 8.
Such results concrete exemplify some of the qualitative arguments 
given in Sec. II.C to explain why the PT should be more efficient 
than the ST around first-order phase transitions.

A different efficiency for the methods is observed not just at the 
phase coexistence, but also for other values of the chemical potential 
$D$ around $D^{*}$.
Figure 9 plots the order parameter $q$ versus $D$ for the PT and ST 
implementations, evaluating the averages at each $M_{a,b} = 10^{4}$ 
MC steps.
Note that overall the PT is already quite close to the values obtained 
from the cluster algorithm, whereas both ST still show some discrepancy, 
specially for $D > D^{*}$.
If now the averages are calculate each $M_{a,b} = 5 \times 10^{4}$ MC 
steps, the ST also becomes closer to the cluster's (inset of Fig. 9). 
Once more such results can be understood in terms of the tunneling 
between the phases. 
For $D  \sim D^{*}$, we still can expect high free energy barriers.
With the ST, the system does not cross such barriers a sufficient
number of times if $M_{a,b} = 10^{4}$.   
By increasing the number of MC steps for the averages, we generate 
a more representative $\Omega$ and thus a better estimation for $m$.

%--------------------- figure -------------------------
%
\begin{figure}
\setlength{\unitlength}{1.0cm}
\includegraphics[scale=0.35]{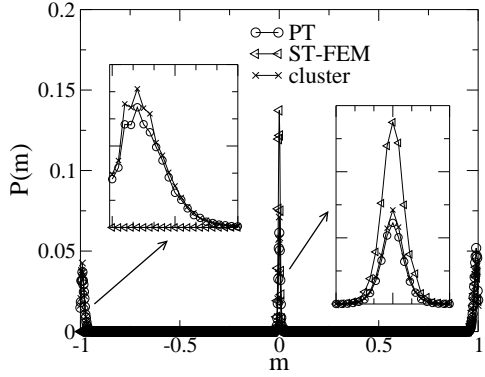}
\caption{Auto-correlation functions versus $\tau$  
from the PT (continuous), ST-FEM (dashed) 
and ST-AM (dotted).}
\end{figure}
%
%--------------------- figure -------------------------                        
%
%
%--------------------- figure -------------------------                        
%
\begin{figure}
\setlength{\unitlength}{1.0cm}
\includegraphics[scale=0.4]{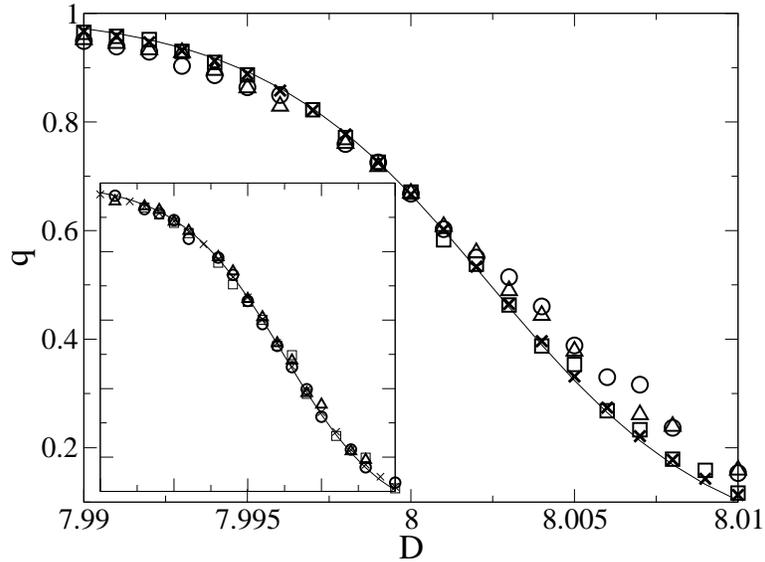}
\caption{Mean probability of exchange versus the temperature 
$T = T_1$ for the PT and ST-FEM.
The symbols $\delta = 1, 2, 3$ refers, respectively, to exchanges 
allowed between first, second and third neighbors (see main text).}
\end{figure}

As a last efficiency measure, we consider the two relevant 
auto-correlation functions, $C_{q}(\tau)$ and $C_{u}(\tau)$, 
shown in Fig. 10.
We should note that although time displaced correlation functions 
are more commonly studied in the context of continuous phase 
transitions, in the present case they are an interesting auxiliary
tool to compare the PT and ST performances.
As it should be, the ST-FEM uncorrelates faster than the ST-AM. 
Nevertheless, we see that the $C$'s decay even faster for the PT 
method (in fact, with a very drastic difference in the case of 
$C_{u}(\tau)$).  

Usually, the frequency (measured in terms of a probability $p^{*}$) 
in which a given tempering method changes the system temperature is 
taken as a good indication of its efficiency.
For the PT and ST algorithms, such quantity respectively reads 
\cite{juan}
$p^{*} = \langle \min \{1,\exp[(\beta_i-\beta_j)({\cal  H}
(\sigma_{i})-{\cal H}(\sigma_{j})] \} \rangle$ 
and 
$p^{*} = \langle \min \{1,\exp[(\beta_{i}-\beta_{j}){\cal H}(\sigma)
+ g_{j}-g_{i}] \}\rangle$.
The averages are over $T_1, \ldots, T_N$, such that 
$p^{*}$ of order $\delta$ is the mean from all the exchanges 
among $T_{n}$ and $T_{n+\delta}$ (see Sec. II.A).

In Fig. 11 we display $p^{*}$ as function of $T = T_1$ 
for the PT and ST-FEM (the ST-AM being similar to the latter), 
with $N = 12$ and $\Delta T = 0.55$.
As it can be seen, for any $\delta$ the ST always presents a 
higher probability of acceptance than the PT, in agreement with 
previous studies \cite{ma,pande2}.
Such findings are in contrast with our results here.
Indeed, larger $p^{*}$'s do not translate into a better performance 
of the ST, at least in the case of phase transitions as argued 
in  Sec. II-C.
Therefore, exchange probabilities alone should be faced with care 
when trying to characterize the best tempering method for a certain 
context.
                 
Finally, we show in Figs. 12 and 13 finite size analysis for the 
total density $q$ and the isothermal susceptibility  
$\chi_{T} = \beta L^{2}(\langle q^{2} \rangle-\langle q  \rangle^{2})$
from the PT and ST-FEM.
Continuous lines correspond to fitting curves by a 
method proposed in Ref. \cite{cluster2}. 
At the phase coexistence, thermodynamic quantities scale 
with the system volume \cite{rBoKo,challa}.
A discontinuous phase transition is characterized by a jump
in the order parameter or even a delta function-like singularity 
for the susceptibility or specific heat.
But this is so only at the thermodynamic limit.
For finite systems not only the order parameter, but also other 
quantities are described by continuous functions 
\cite{fiore8,wang,cluster2}. 
We should emphasizes that smooth curves are obtained only when one 
uses a simulation dynamics which correctly yields an appropriate
sampling.
For instance, from simple Metropolis algorithms, neither the crossing 
among isotherms nor accurate finite size analysis for smooth curves 
are possible.
It is due to the presence of hysteresis effects 
\cite{fiore8,cluster1,cluster2}, which hence demand tempering 
enhanced algorithm.      
From the plots we see that both the PT and ST give fairly good results.
However, the cluster continuous curve \cite{cluster2} is smoother
and better fitted in the PT case, specially for the larger $L = 30$
value.

%
%--------------------- figure -------------------------                        
%
\begin{widetext}
\begin{center}
\begin{figure}
\setlength{\unitlength}{1.0cm}
\includegraphics[scale=0.4]{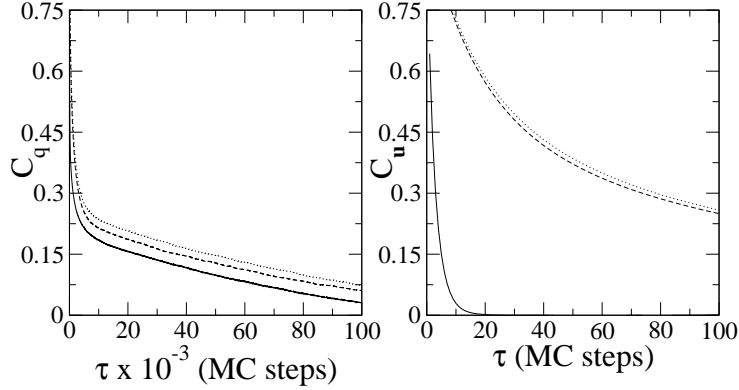}
\caption{$q$ versus $D$ for $L$ equal to 10 (circle), 20 (square) 
and 30 (triangle), calculated from the (a) PT and (b) ST-FEM. 
Continuous lines are fitting results \cite{fiore8,cluster2}.
The curves collapse if plotted as 
$q \times (D-D^{*}) L^2$ (insets).}
\end{figure}
%
%
%--------------------- figure -------------------------                        
%
\begin{figure}
\setlength{\unitlength}{1.0cm}
\includegraphics[scale=0.42]{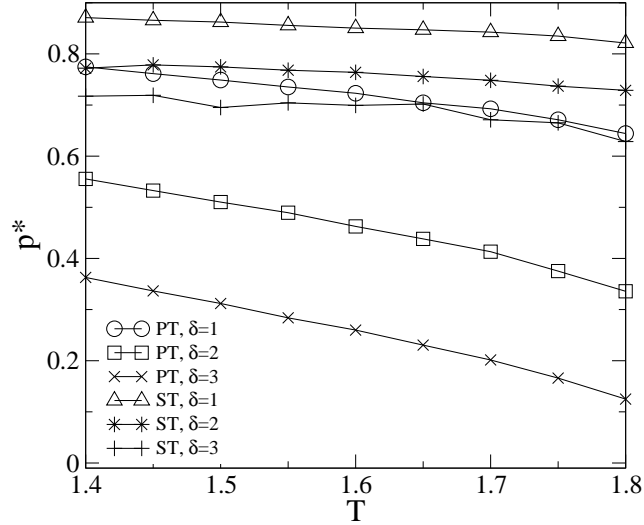}
\caption{Susceptibility versus $D$ for $L$ equal to 10 (circle), 
20 (square) and 30 (triangle), calculated from the (a) PT and (b) 
ST-FEM. 
The curves collapse if plotted as 
$\chi_{T}/L^{2} \times (D-D^{*})L^{2}$ (insets).}
\end{figure}
\end{center}
\end{widetext}
%
%--------------------- figure -------------------------                        
%

\section{Remarks and Conclusion}

In this paper we have presented a comparative study between two 
important enhanced sampling methods, namely, simulated (ST) 
and parallel (PT) tempering, considering spin-lattice models
at phase transition conditions.
Special attention has been payed to first-order phase transitions 
at low temperatures (for the BEG model).
In such regimes, more standard algorithms often give poor results 
because their difficulties to overcome the large free-energy 
barriers in the phase space, leading, e.g., to ergodicity 
breaking and artificial algorithm-induced hysteresis.
We also have investigated the less critical case of second 
order-phase transition -- for which no free-energy barriers 
exist but there is the formation of strongly correlated
clusters (basin regions) \cite{landau-binder} -- for the well 
understood Ising model. 

As for the tempering implementations, we have followed the usual 
PT procedure, but allowing temperature exchanges between 
non-adjacent replicas. 
For the temperature change probability weights $g$ in the ST, we 
have assumed a recent proposed approximation \cite{pande} (ST-AM) 
and a new alternative exact approach (ST-FEM), based on the 
eigenvalues of the transfer matrix \cite{sauerwein}.
The ST-FEM here is formally similar to that in Ref. \cite{ma}, 
but avoids the necessity to implement more complicated recursive 
procedures to estimate the partition function.

Different comparative analysis, both at the transient regime 
and already at the steady state, have been carried out.
Despite the facts that: 
(i) after long times (thus demanding large computational effort)
the final results from the PT and ST are similar;
and (ii) the PT displays a smaller exchange probability than 
the ST; 
we have found that for discontinuous phase transitions the PT is 
always more efficient in any verified aspect.
The main reason for this is basically that the PT enables the 
system to cross free-energy barriers more frequently than the 
ST: either at or near phase coexistence conditions (as explicit
illustrated, e.g., in Figs. (7) and (8)).
Furthermore, besides the quantitative numerical results, we also 
have presented heuristic arguments for why it should be expected.

Results for the instructive Ising model at the critical temperature 
(second-order phase transition) have also agreed with our qualitative 
predictions.
Indeed, far away from $T_c$ it has been reported a faster convergence 
for the ST \cite{ma}. 
We have shown that for $T \sim T_c$ just the opposite takes 
place, with the auto-correlations decaying faster for the PT. 

For completeness, we also have analyzed other values of $K/J$ for
the BEG model (not shown), in particular for $K/J=0$, the so
called Blume-Capel model.
The calculations at the first-order transition 
($T_1 = 0.4$ and $D=1.9968$) have corroborated the higher 
efficiency of the PT over the ST.
More specifically, until $M_{tot} = 3 \times 10^{7}$, the system 
when simulated with the ST-AM has not reached the steady state, 
whose values for the thermodynamic quantities were different 
from those obtained by the ST-FEM, PT and cluster algorithms.
Furthermore, the ST-FEM have agreed with the PT and cluster only 
for long $M_{tot}$'s.
Time-displaced correlation functions decays and actual 
thermodynamic quantities convergence were always faster for the PT.

A second contribution of this work has been an (numerically 
simpler) alternative way to calculate the exact $g$ in the ST method.
When comparing the ST-AM with the ST-FEM, we have found that the 
ST-FEM allows the system to converge to steady regime quicker 
than the ST-AM (see above). 
In addition, at the steady state, configurations generated by 
ST-FEM uncorrelate faster than those by the ST-AM.
On the other hand, with respect to the frequency in which the 
system tunnels between different phases at the coexistence
and the final sough thermodynamic quantities, both implementations 
are similar, but the latter only for long $M_{tot}$'s.

Summarizing, at phase transition regimes the PT and ST provide 
the same results for long (sometimes even costly) simulations. 
However, we find that for all the tested measures, the parallel 
converges faster than the simulated tempering.
Also, even in such situation of a better performance from the PT, 
still the rate of temperature switching is higher for the ST. 
Thus, another message from our work is that alone, the switching 
rates are not sufficient to characterize the efficiency of a 
tempering enhanced sampling algorithm.

\section*{Acknowledgements}

We acknowledge researcher grants by CNPq.
Financial support is also provided by CNPq-Edital Universal, 
Funda\c c\~ao Arauc\'aria and Finep/CT-Infra.

%%%%%%%%%%%%%%%%%%%%%%%%%%%%%%%%%%%%%%%%%%%%%%%%%%%%%%%%%%%%%%%%%%%%%%%%%%  

\end{document}